\RequirePackage{fix-cm}
\documentclass[smallcondensed]{svjour3}     
\smartqed  
\usepackage{graphicx}
\usepackage{footnote}

\begin{document}

\title{Remarks on Chebyshev representation of ephemeris 
}


\author{Yanning FU
}


\institute{Yanning FU \at
              Purple Mountain Observatory \\
              Tel.: +81-25-83332137\\
              Fax: +81-25-83332138\\
              \email{fyn@pmo.ac.cn}           
}

\date{Received: date / Accepted: date}

\maketitle

\begin{abstract}
Chebyshev coefficients of a coordinate Chebyshev representation can be used to form representations of velocity. One way is to directly apply them to the derivatives of Chebyshev polynomials, another is to compute from them the Chebyshev coefficients of velocity Chebyshev representation. The advantages of the latter over the former ways are illustrated. Also, the approach of generating Chebyshev coefficients developed by Newhall \cite{New89} is extended such that coordinate, velocity and acceleration are consistently treated.
\end{abstract}

\section{Introduction}
\label{intro}
Chebyshev expansions, i.e. linear combinations of Chebyshev polynomials, are commonly used to represent high-precision numerical ephemerides of celestial bodies, e.g. the well-known DE, EPM and INPOP planetary and lunar ephemerides. Newhall \cite{New89} develops an approach of generating Chebyshev representations of coordinates. Such a representation, the combination coefficients of which are widely distributed and used, is piecewise but continuous up to its first derivative (i.e. the so-called coordinate velocity), e.g. \cite{GLFM}. When applied to the first derivatives of Chebyshev polynomials, the same coefficients are used to compute the velocity.

Here are some simple facts about the above usual practice. Firstly, velocities are not computed using Chebyshev expansions. The advantages of Chebyshev expansion thus become irrelevant. And, as will be obvious later, there are often cases when certain amount of computation can be saved without increasing the size of the distributed ephemeris file. Secondly, the maximum error of a velocity representation, like that of a coordinate one, can be estimated by using the distributed Chebyshev coefficients. Thirdly, though not realized in the standard interpolation softwares, accelerations can be computed in the same way as velocities. This is of interest because it allows direct comparison between the actually realized force models underlying different ephemerides by using publicly available ephemeris data.

Sect.~\ref{sec:2} recalls the basic knowledge of Chebyshev expansions necessary for detailing the brief remarks made in the previous paragraph. And the detailed remarks are presented in Sect.~\ref{sec:3}. Sect.~\ref{sec:4} is devoted to extending the approach developed in \cite{New89}, so that coordinate, velocity and acceleration are consistently treated.

\section{Chebyshev expansion}
\label{sec:2}
A Chebyshev expansion of degree $N$ writes
\begin{equation}\label{eqn ChExpansion}
p(x)=\sum_{n=0}^{N}p_n{T_n(x)}
\end{equation}
where $x$ and $p_n$'s are referred to as Chebyshev time and Chebyshev coefficients, respectively, and the Chebyshev polynomial $T_n(x)$ is a polynomial of degree $n$ defined on $x\in[-1,1]$ and recursively given by
\begin{equation}\label{eqn ChPolynomials}
\begin{array}{l}
T_0(x)=1, \ T_1(x)= x, \\
T_{n}(x)=2xT_{n-1}(x)-T_{n-2}(x)\ \ \ (n=2,...,N).
\end{array}
\end{equation}
For all non-negative integers $n$, we have
\begin{equation}\label{eqn supT}
\sup_{x\in[-1,1]}|T_n(x)|=1.
\end{equation}

The first derivative of $p$ with respect to $x$ writes
\begin{eqnarray}
v(x)=\sum_{n=0}^{N}p_n{T_n'(x)}&=&\sum_{n=1}^{N}p_n{T_n'(x)}\label{eqn DChExpansion 1} \\
                               &=&\sum_{n=0}^{N-1}v_n{T_n(x)},\label{eqn DChExpansion 2}
\end{eqnarray}
where
\begin{equation}\label{eqn DChPolynomials}
\begin{array}{l}
 T_0'(x)=0,\ T_1'(x)=1, \\
 T_{n}'(x)=2xT_{n-1}'(x)+2T_{n-1}(x)-T_{n-2}'(x)\ \ \ (n=2,...,N),
\end{array}
\end{equation}
and, with $v_{N}=0$,
\begin{equation}\label{eqn DChCoefficients}
\begin{array}{l}
v_{N-1}=2Np_N, \\
v_{n}=2(n+1)p_{n+1}+v_{n+2}\ \ \ (n=N-2,...,1), \\
v_0=p_{1}+v_{2}/2.
\end{array}
\end{equation}
For $n\geq 0$, we have\footnote{Less known than (\ref{eqn supT}), this conclusion can be proved by using another equivalent definition of Chebyshev polynomial, i.e. $T_n(x)=\cos(n \mathrm{Arccos}(x))$}.
\begin{equation}\label{eqn supdT}
\sup_{x\in[-1,1]}|T'_n(x)|=n^2.
\end{equation}

The second derivative of $p$ with respect to $x$ writes
\begin{eqnarray}
a(x)=\sum_{n=0}^{N}p_n{T_n''(x)}&=&\sum_{n=2}^{N}p_n{T_n''(x)}\label{eqn DDChExpansion 1}\\
    &=&\sum_{n=0}^{N-2}a_n{T_n(x)}\label{eqn DDChExpansion 2},
\end{eqnarray}
where
\begin{equation}\label{eqn DDChPolynomials}
\begin{array}{l}
 T_0''(x)=0, \\
 T_1''(x)=0,\ T_2''(x)=4, \\
 T_{n}''(x)=2xT_{n-1}''(x)+4T_{n-1}'(x)-T_{n-2}''(x)\ \ \ (n=3,...,N),
\end{array}
\end{equation}
and, with $a_{N-1}=0$,
\begin{equation}\label{eqn DDChCoefficients}
\begin{array}{l}
a_{N-2}=2(N-1)v_{N-1}, \\
a_{n}=2(n+1)v_{n+1}+a_{n+2}\ \ \ (n=N-3,...,1), \\
a_0=v_{1}+a_{2}/2.
\end{array}
\end{equation}

\section{Chebyshev representation of ephemeris}
\label{sec:3}
A Chebyshev representation of ephemeris of a celestial body allows users to compute the body's dynamical state (coordinate, velocity and acceleration) by using a set of piecewise Chebyshev expansions. In practice, this is economically achieved by only providing Chebyshev coefficients of coordinate representaions. To be specific, let's focus on a particular Chebyshev expansion representing a coordinate and being valid on a granule $[t_b,t_e]$ of the ephemeris time $t$. Introducing the Chebyshev time \begin{equation}\label{eqn ChTime}
x=-1+2\frac{t-t_b}{t_e-t_b}\in[-1,1],
\end{equation}
one computes the coordinate, velocity and acceleration by using (\ref{eqn ChExpansion}), (\ref{eqn DChExpansion 1}) and (\ref{eqn DDChExpansion 1}), respectively\footnote{Back to $t$, the velocity writes $\frac{dp}{dt}=\frac{2}{t_e-t_b}v$ and the acceleration $\frac{d^2p}{dt^2}=(\frac{2}{t_e-t_b})^2a$.}. To do so, Chebyshev polynomials and their respective first and second derivatives have to be computed for each required value of $x$.

The dynamical state can also be computed using (\ref{eqn ChExpansion}), (\ref{eqn DChExpansion 2}) and (\ref{eqn DDChExpansion 2}). An advantage of doing so is time-saving. This is because the $x$-dependant parts, i.e. the Chebyshev polynomials, are common to these three expansions, while the $x$-independent parts, i.e. the Chebyshev coefficients, need only to be computed once for all $x\in[-1,1]$.

As representations in the form of Chebyshev expansion, (\ref{eqn DChExpansion 2}) and (\ref{eqn DDChExpansion 2}) allow one to estimate the maximum representation error of velocity and acceleration, respectively, in a way by which \cite{New89} estimates that of coordinate. For this, a necessary assumption for $n\geq N$ is either $p_n\sim 0$ or $|p_{n+1}|/|p_{n}| < \epsilon$, where $\epsilon$ is a small positive constant. This assumption is true with $\epsilon \sim 0.1$ in the case of distributed planetary and lunar representations, for which the length of granule in $t$ and the degree $N$ of representation are conveniently chosen \cite{New89}. Together with $|T_n(x)|\leq 1$, the above assumed condition implies that the coordinate representation error is bounded from above by
\begin{equation}\label{eqn perr}
\delta_p=\sum_{n=N+1}^\infty |p_n|\leq |p_{N+1}|\sum_{i=0}^\infty\epsilon^i=\frac{|p_{N+1}|}{1-\epsilon}\leq\frac{\epsilon |p_N|}{1-\epsilon} \sim \epsilon |p_N|.
\end{equation}
Checking with an Inpop10e ephemeris file\footnote{Namely, the ephemeris file inpop10e{\textunderscore}TDB{\textunderscore}m1000{\textunderscore}p1000{\textunderscore}littleendian.dat available at http://www.imcce.fr/inpop/.} valid on a time span of 2000 years, we find that $|v_{N}|/|v_{N-1}|<0.125$, provided that $|p_{N+1}|/|p_{N}|<\epsilon=0.1$. This is understandable since $v_{n} \sim 2(n+1)p_{n+1}$, which means that
\begin{equation}\label{eqn vratio}
\frac{|v_{n}|}{|v_{n-1}|} \sim \frac{n+1}{n} \frac{|p_{n+1}|}{|p_n|} < \epsilon (1+\frac{1}{n})
\left\{
\begin{array}{lr}
\sim 1.2\epsilon \leq 0.12 & (n=\min(N)=5), \\
\sim \epsilon=0.1 & (n\ \mathrm{large}).
\end{array}
\right.
\end{equation}
From (\ref{eqn vratio}), the same argument as in the case of coordinate leads to the following apparent estimation of the maximum error of the velocity representation
\begin{equation}\label{eqn verr}
\delta_v=\epsilon |v_{N-1}| \sim 2N \epsilon |p_N| \sim 2N\delta_p.
\end{equation}
Similarly, the maximum error of the acceleration representation is estimated as
\begin{equation}\label{eqn aerr}
\delta_a=\epsilon |a_{N-2}| \sim 2(N-1) \delta_v \sim 4N(N-1)\delta_p.
\end{equation}
Given sampling data of a function of time, the precision of the best-fit interpolation polynomial depends on its degree, as well as the behavior of the function in the considered time interval. For each given body, \cite{New89} chooses the degree of the polynomials representing coordinates, $N$ appearing in (\ref{eqn ChExpansion}), and the length of granule, $L=t_e-t_b$ corresponding to 2 Chebyshev time unit $\mathrm{u_x}$, such that $\delta_p \sim 0.5\mathrm{mm}$.\footnote{The error in 3-dimensional position is then at the level of sub-millimeter, which is smaller than that of the present-day laser ranging measurement.} Table \ref{tab va error} lists the corresponding maximum errors of velocity and acceleration representations, as calculated using (\ref{eqn verr}) and (\ref{eqn aerr}), respectively.

\begin{table}
\caption{Estimated maximum errors of velocity and acceleration representations. The estimated values in Chebyshev time unit $\mathrm{u}_x$ are computed using (\ref{eqn verr}) and (\ref{eqn aerr}), where the maximum error of coordinate representation $\delta_p=0.5\mathrm{mm}$ \cite{New89}. The corresponding values in ephemeris time $t$ also depend on the ephemeris time length $L$ of granule.}
\label{tab va error}
\center
\begin{tabular}{ccccc}
\hline
  Body & $L$ & $N$ & $\delta_v=2N\delta_p$ & $\delta_a=4N(N-1)\delta_p$  \\
       & day &     & $\mathrm{mm/day\ [mm/u_x]} $ & $\mathrm{mm/day^2\ [mm/u_x^2]}$  \\
\hline
   Mercury    & $\ 8 $ & $ 13 $ & $ 3.2\ \ [  13.0 ]$ & $  19.5 \ \  [ 312.0  ]$ \\
   Venus      & $ 16 $ & $\ 9 $ & $ 1.1\ \ [ \ 9.0 ]$ & $  \ 2.2 \ \ [ 144.0  ]$ \\
   Earth-Moon & $ 16 $ & $ 12 $ & $ 1.5\ \ [  12.0 ]$ & $  \ 4.1 \ \ [ 264.0  ]$ \\
   Mars       & $ 32 $ & $ 10 $ & $ 0.6\ \ [  10.0 ]$ & $  \ 0.7 \ \ [ 180.0  ]$ \\
   Jupiter    & $ 32 $ & $\ 7 $ & $ 0.4\ \ [ \ 7.0 ]$ & $  \ 0.3 \ \ [\ 84.0  ]$ \\
   Saturn     & $ 32 $ & $\ 6 $ & $ 0.4\ \ [ \ 6.0 ]$ & $  \ 0.2 \ \ [\ 60.0  ]$ \\
   Uranus     & $ 32 $ & $\ 5 $ & $ 0.3\ \ [ \ 5.0 ]$ & $  \ 0.2 \ \ [\ 40.0  ]$ \\
   Neptune    & $ 32 $ & $\ 5 $ & $ 0.3\ \ [ \ 5.0 ]$ & $  \ 0.2 \ \ [\ 40.0  ]$ \\
   Pluto      & $ 32 $ & $\ 5 $ & $ 0.3\ \ [ \ 5.0 ]$ & $  \ 0.2 \ \ [\ 40.0  ]$ \\
   Moon       & $\ 4 $ & $ 12 $ & $ 6.0\ \ [  12.0 ]$ & $   66.0 \ \ [ 264.0  ]$ \\
   Sun        & $ 16 $ & $ 10 $ & $ 1.2\ \ [  10.0 ]$ & $  \ 2.8 \ \ [ 180.0  ]$ \\
\hline
\end{tabular}
\end{table}

From Table \ref{tab va error}, we know that the velocity representation errors are extremely small as compared with the most precise doppler velocity measurement used in the present-day spacecraft tracking (several millimeters per second). Therefore, there should be many cases when larger errors are tolerable. In these cases, truncated velocity representations may be used providing that their respective maximum error is known. In the context of truncation, the well known advantage of (\ref{eqn DChExpansion 2}) over (\ref{eqn DChExpansion 1}) can be quantified. For example, neglecting the last term of (\ref{eqn DChExpansion 2}) will induce an error no more than
$|v_{N-1}T_{N-1}(x)|\leq |v_{N-1}|=2N|p_N|$, while the error induced by neglecting the last term of (\ref{eqn DChExpansion 1}) can be as large as
$|p_NT'_N(x)|\leq N^2 |p_N|$.

\section{Chebyshev coefficient generation}
\label{sec:4}
In \cite{New89}, an efficient approach is developed to generate the coefficients of a piecewise Chebyshev coordinate representation. What is special of this approach is twofold. Firstly, it uses not only sampling data of coordinate but also of velocity to form a generally overdetermined equation system for the coefficients. Secondly, it uses exact constraints at both boundary time instants of each representation piece. The advantage of doing so is also twofold: (1) The effect of noises on the derivative of the coordinate representation, used as a velocity representation, is reduced; (2) Both coordinate and velocity representations are continuous.

There is a price to pay for the first advantage. Indeed, the noises induced by the velocity equations should have negative effects on the coordinate representation. In order to balance between the advantage to gain and the price to pay, \cite{New89} uses a universal weighting scheme based on experiments. In this scheme, velocity equations are weighted at 0.4 relative to coordinate ones. Though the scheme may be oversimplified, its ephemeris-independent nature allows the approach to express the searched Chebyshev coefficients as linear combinations of sampling data with ephemeris-independent combination coefficients. This should be considered as another strong point of the approach, because these combination coefficients, depending only on the degree of the coordinate representation and the set of sampling time instants \cite{New89}, apply to any ephemeris.

In a broad sense, how to assign weights to sampling equations may be considered as a matter of choice. To make a decision, one needs a pre-determined criterion, which can be different for different purposes. Here are two extreme examples: If velocity is not required, then it's better not to use velocity equations, i.e. they should be assigned with zero weight; And, if acceleration is required, then it's better to use also acceleration equations, which may be weighted in a way similar to velocity equations, namely at 0.4 relative to velocity equations. Another reasonable and ephemeris-independent weighting scheme is based on the maximum-representation-error relation $\delta_p:\delta_v:\delta_a=1:2N:4N(N-1)$ (see section \ref{sec:3}). For this we remark that, contrary to sampling by observations, here the sampling data are assumed to be exact, while the noises come from the non-exact representation models. It is then interesting to present an extended approach without specifying a particular weighting scheme.

Let $\{P_m,V_m,A_m\}$, $m=1,...,M$, be the instantaneous dynamical states at $x_m \in (-1,1)$, respectively, which are part of the data carried by a numerical integration program \cite{New89}. The extended objective function writes
\begin{equation}\label{eqn objfun}
\chi^2(p_0,...,p_N)=\sum_{m=1}^{M} [w_p(p(x_m)-P_m)]^2+[w_v(v(x_m)-V_m)]^2+[w_a(a(x_m)-A_m)]^2,
\end{equation}
where $p,v$ and $a$ are represented as (\ref{eqn ChExpansion}), (\ref{eqn DChExpansion 1}) and (\ref{eqn DDChExpansion 1}), respectively, and the weights $w_p:w_v:w_a$ are assumed to be ephemeris-independent.

Let $\{P_{bl},V_{bl},A_{bl}\}$, $l=1,2$, be two additional dynamical states sampled at $x=x_{b1}=-1$ and $x_{b2}=1$, respectively. The imposed equality constraints are
\begin{equation}\label{eqn constraint}
\left\{
\begin{array}{lcr}
b_{pl}=p(x_{bl})-P_{bl}=0, & \\
b_{vl}=v(x_{bl})-V_{bl}=0, & \ \ \ \ (l=1,2).\\
b_{al}=a(x_{bl})-A_{bi}=0, &
\end{array}
\right.
\end{equation}

 Because $p,v$ and $a$ are all linear in the searched Chebyshev coefficients $p_0,..., p_N$, our optimization problem with equality constraints is linear. The unique solution to this problem can be found by the well-known method of Lagrange multiplier. Let $\lambda_{pl},\lambda_{vl},\lambda_{al}\ (l=1,2)$ be the Lagrange multipliers and
\begin{equation}\label{eqn Lobjfun}
L(p_0,...,p_N,\lambda_1,...,\lambda_6)=\chi^2+\sum_{l=1}^{2}[\lambda_{pl}b_{pl}+\lambda_{vl}b_{vl}+\lambda_{al}b_{al}]
\end{equation}
the Lagrange function. The Chebyshev coefficients, together with the 6 multipliers, minimizes $L$ and solves the normal equations
\begin{equation}\label{eqn normal}
\left\{
\begin{array}{lcr}
\partial L/\partial p_n=0 & & (n=0,...,N), \\
\partial L/\partial \lambda_{pl}=0,\ \partial L/\partial \lambda_{vl}=0,\ \partial L/\partial \lambda_{al}=0 & & (l=1,2).
\end{array}
\right.
\end{equation}
To present explicitly this equation system in matrix form, we introduce three groups of matrices, namely: The parameters to be optimized
\begin{equation} \label{eqn matrices p2bo}
\begin{array}{lcr}
s=[p_0,...,p_N]^t & & (\mathrm{order}\ (N+1) \times 1),\\
\lambda=[\lambda_{p1},\lambda_{p2},\lambda_{v1},\lambda_{v2},\lambda_{a1},\lambda_{a2}]^t & & (\mathrm{order}\ 6 \times 1),
\end{array}
\end{equation}
where and in the following the superscript $t$ stands for transposition; The dynamical state variables sampled at inner and boundary time instants
\begin{equation}\label{eqn matrices S}
\begin{array}{lcr}
S=[P_1,...,P_M,V_1,...,V_M,A_1,...,A_M]^t & & (\mathrm{order}\ 3M \times 1), \\
S_b=[P_{b1},P_{b2},V_{b1},V_{b2},A_{b1},A_{b2}]^t & &  (\mathrm{order}\ 6 \times 1);
\end{array}
\end{equation}
And the Chebyshev polynomials and their first and second derivatives evaluated at inner and boundary instants
\begin{equation}\label{eqn matrices T}
\begin{array}{lcr}
T=[T_n(x_m)],\ T'=[T'_n(x_m)],\ T''=[T''_n(x_m)] & & (\mathrm{order}\ (N+1) \times M), \\
T_b=[T_n(x_{bl})],\ T_b'=[T'_n(x_{bl})],\ T_b''=[T''_n(x_{bl})] & & (\mathrm{order}\ (N+1) \times 2).
\end{array}
\end{equation}
As building blocks, the third group matrices form the following matrices,
\begin{equation}\label{eqn matrices T}
\begin{array}{lcr}
F=F^t=2(w_p^2 TT^t+w_v^2 T'T'^t+w_a^2 T''T''^t) & & (\mathrm{order}\ (N+1) \times (N+1)), \\
G=2(w_p^2 T,w_v^2 T',w_a^2 T'') & & (\mathrm{order}\ (N+1) \times 3M), \\
H=(T_b,T'_b,T''_b) & & (\mathrm{order}\ (N+1) \times 6). \\
\end{array}
\end{equation}
Straightforward calculation leads us to the following matrix form of (\ref{eqn normal})
\begin{equation}\label{eqn Mequation}
\mathcal{M}_L
\left(
\begin{array}{c}
s \\
\lambda
\end{array}
\right)
=
\mathcal{M}_R
\left(
\begin{array}{l}
S \\
S_b
\end{array}
\right),
\end{equation}
where, with $0_{m \times n}$ and $I_{m \times n}$ standing respectively for zero and unit matrices of order $m \times n$,
\begin{equation}\label{eqn FGH}
\begin{array}{lcr}
\mathcal{M}_L=\left(
\begin{array}{cc}
F & H \\
H^t & 0_{6 \times 6}
\end{array}
\right) & &(\mathrm{order}\ (N+7) \times (N+7)), \\
\mathcal{M}_R=
\left(
\begin{array}{cc}
G & 0_{(N+1) \times 6} \\
0_{6 \times 3M} & I_{6 \times 6}
\end{array}
\right) & &(\mathrm{order}\ (N+7) \times (3M+6)).
\end{array}
\end{equation}
Scaling weight-dependent matrices, $F$ and $G$, with the same non-zero constant factor $\alpha$, one obtains a new matrix equation, which can be derived from the objective function $\alpha\chi^2$ having the same conditional extreme point as our $\chi^2$. To be explicit, we remark that, if $[s,\lambda]^t$ solves (\ref{eqn Mequation}), then $[s,\lambda/\alpha]^t$ solves the new equation. This means that, as expected, $w_p,w_v$ and $w_a$ can be simultaneously scaled.

The solution of (\ref{eqn Mequation}) can be written as $[s,\lambda]^t=\mathcal{M}[S,S_b]^t$, where $\mathcal{M}=\mathcal{M}_L^{-1}\mathcal{M}_R$ is a matrix of order $(N+7) \times (3M+6)$. Let $\mathcal{M}_n$ be the $n$th row vector of $\mathcal{M}$, then the searched Chebyshev coefficients write
\begin{equation}\label{eqn solution s}
\begin{array}{lcr}
p_n=\mathcal{M}_{n+1}\left(
\begin{array}{l}
S \\
S_b
\end{array}
\right) & & (n=0,...,N).
\end{array}
\end{equation}



\end{document}